\documentclass[12pt,a4paper]{article}
\usepackage{cite}
\usepackage[dvips]{epsfig}

\renewcommand{\thefootnote}{\fnsymbol{footnote}}

\begin{document}

\begin{center}
 {\large \bf  Relativistic Phase Space: Dimensional Recurrences}
\end{center}
 \vspace{1cm}

\begin{center}
 R.~Delbourgo\footnote{Email address: Bob.Delbourgo@utas.edu.au} 
 and M.L.~Roberts\footnote{Email address: Martin.Roberts@utas.edu.au}
\\
 \vspace{1cm}
{\em School of Mathematics and Physics,
University of Tasmania,\\ GPO Box 252-21, Hobart,
Tasmania 7001, Australia}
\end{center}


\begin{abstract}
We derive recurrence relations between phase space expressions in
different dimensions by confining some of the coordinates to tori
or spheres of radius $R$ and taking the limit as $R \rightarrow \infty$.
These relations take the form of mass integrals, associated with extraneous
momenta (relative to the lower dimension), and produce the result in the 
higher dimension.
\end{abstract}

\newpage
\renewcommand{\thefootnote}{\arabic{footnote}}
\setcounter{footnote}{0}

\section{Establishing the Recurrences} 

Formulations of field theories in higher dimensions are now quite 
commonplace, with 10 to 12 dimensions featuring prominently, especially in the 
context of M-theory. Mostly the coordinates which are additional to the usual 
four space-time ones are either very tiny or else the fields possess a 
severely damped behaviour as one moves away from the 3-brane. Thus the extra 
coordinates are characterized by one or more length scales $R$, that are
generally miniscule or, if larger, only affect gravity. At the other extreme
one can contemplate the $R$ as being huge; indeed the method of `box 
normalization' with a large $R$ has a venerable pedigree and allows us to
describe vacuum diagrams or compute properties per unit volume when they so
depend. The same method also permits us to make sense of quantities like the 
volume of space-time or $\delta^4(0)$ in the limit as $R \rightarrow \infty$.

One of the primary objects of interest in these higher dimensions $D\equiv 
2\ell$ is the behaviour of relativistic $N$-body phase space $\rho^D_N$ since 
it primarily governs the magnitudes of transmutations, ignoring amplitude
modulation. Based on earlier coordinate space methods \cite{ADM}, 
this behaviour has been recently studied \cite{GP} and codified 
\cite{BDR}; a summary of how the $N$-body result in {\em fixed}, flat $D$ 
space can be evaluated by means of Almgren recurrences \cite{Alm}---mass 
integrals over smaller $N$ but with the {\em same} $D$--- was given in ref.[2]. 
In this paper we wish to establish 
relations between $\rho$ having the same $N$, but different $D$, which are
quite interesting in their own right. They may well have some use in the
context of recent developments in string or M-theory or $p$-brane physics;
for such theories possess a set of length scales $R_i$ (or parameters
arising in the extended metric) which serve to
constrain the motion of particles to subspaces or `branes' of lower dimension.
Each theory produces its own particle spectrum whose spacing is determined
by the $R_i$. As the limit $R_i\rightarrow\infty$ is taken, we may
anticipate that the particles freely explore the higher-dimensional space,
and the corresponding phase space at a given energy reduces to the
relativistic phase space for the entire `bulk'.

The strategy for deriving such recurrence relations in the flat space
limit is quite simple. We
just confine one or several of the $D$-dimensional coordinates to a 
torus or sphere of radius $R$ and take the limit $R\rightarrow \infty$ at the
very end, in much the same way that box normalization is handled. The act of
confinement creates a series of discrete modes in the restricted coordinates 
and phase space must be summed over the various modes, subject to mass
bounds. By choosing the topology of the extra space to be spherical or
toroidal, the masses of the modes are easy to work out. (Had we been
considering more complicated topologies, the mass spectra would have been
much harder to calculate although we still expect the $R_i\rightarrow\infty$ 
limit to yield results coinciding with our choices.)
Because the contributions to the masses from the confined coordinates
are inversely proportional to $R$, the summation reduces to a mass integral in
the limit of enormous $R$: hence the result $\rho_N^D$ in higher $D$ is 
expressible as a set of mass integrals of $\rho$ over lower $D$, but with the 
same $N$; one can readily understand this as the effect of integrating over 
extra momentum components relative to those in the lower dimension. The
forms of such connections are rather intriguing and some are not at all
obvious; in fact for larger $N$ they are quite intricate. We know of course
in advance that they must work out somehow; the interest is in the `somehow',
not why they do so. For smaller values of $N$ we are able to check their
validity, but verifying them for $N>3$ is a daunting task in many cases. 

In the following section we suppose that one or several of the coordinates are
confined to a torus, so the recurrence relation is effectively that between
$\rho^D$ and $\rho^{D-1}$. Its nature is pretty simple since there 
is only one extra coordinate to contend with, so we are dealing with a 
one-dimensional sum or integral in the continuum limit. We show how the 
recurrence pans out for few-body processes. The next two sections deal with 
the case when there are several extra coordinates (confined to a sphere) and 
here we encounter a multidimensional summation or integration, which is 
nontrivial. When $N=2$ we demonstrate how the relations work out for any 
number of extra coordinates, but for $N=3$ we have only succeeded in following 
through the connection when the dimension difference equals two or more, 
though no doubt it must apply to any $N$ value.

\section{Relations between $D$ and $D-1$ phase space}

Let us begin by specifying our notation. Our metric is +,-,-,-... with a total
of $D$ coordinates. The $N$-body phase space integral in {\em flat} space is 
defined by
\begin{equation}
 \rho_N^D = \left(\prod_{i=1}^N\int d\Omega_{p_i}\right)(2\pi)^D\delta^D
 (p-\sum_{i=1}^N p_i)
\end{equation}
where $d\Omega_{p_i}\equiv\theta(p_{i0})\delta(p_i^2-m_i^2)d^Dp_i/(2\pi)^{D-1}$.
We separate spacetime coordinates $x$ into $(D-n)$ coordinates called $\bf{x}$ 
and extra ones labelled $\vec{y}$, $n$ of which are independent; 
likewise for the conjugate momenta. (For the purpose of this section $n$ 
equals 1, but we shall consider other $n$ values later on.) 
The general aim of the exercise is to establish a connection 
between $\rho_N^D$ and $\rho_N^{D-n}$ and see how that works out analytically 
because it is nontrivial for large $N$ or $n$.

The first step is to confine (periodic) $y$ to the circumference of a circle 
of radius $R$. Thus the space is considered to have the direct product topology
$M^{D-1}(x)\otimes T^1(y)$. Then 
Fourier expand a (real) field $\Phi$ into modes in the standard way so as to 
fix the normalization correctly:
\begin{equation}
 \Phi({\bf{x}},y)=\sum_{k=-\infty}^\infty \phi_k({\bf{x}})
 \exp(iky/R)/\sqrt{2\pi R}; \quad \phi_k^*({\bf{x}}) = \phi_{-k}({\bf{x}}).
\end{equation}
Being complex in general, $\phi_k^*$ can be regarded as the antiparticle field
to $\phi_k$ where $k$ is positive say. $y$ must run from $-\pi R$ to $\pi R$, 
to ensure that, upon integration over $y$, the free action takes its proper
form,
\begin{equation}
 S_{\rm free}= \frac{1}{2}\int\!\! d^D\!x[(\partial_x\Phi)^2 - m^2\Phi^2]
 =\sum_k \frac{1}{2}\int\!\!d^{D-1}\!{\bf x}[(\partial_{\bf x}\phi_k^*)
 (\partial_{\bf x}\phi_k) - m_k^2\phi_k^*\phi_k],
\end{equation}
where $m_k^2 \equiv m^2 + (k/R)^2$ corresponds to the mode $k$ mass squared.
Note that if one restricts the sum to positive $k$ (because of repetition) the 
factor of 1/2 disappears and one gets the right normalization for a complex 
field. A zero-mode field which is of course real and $y$-independent is 
given by $\Sigma({\bf x}) = \sigma({\bf x})/\sqrt{2\pi R}$.

To determine the phase space integral we consider the point interaction between
a heavy field $\Sigma$ (with mass $m_0$) decaying into $N$ distinguishable 
fields which carry their own distinct quantum numbers, so $\Sigma$ matches all 
of them. (One may also consider the case where some of the final particles are 
identical, but that just serves to introduce symmetry factors which must be 
taken into account and adds little to the discussion.) Write the effective
interaction as ${\cal L}=\phi_1\ldots\phi_N\Sigma$, so that integration over
the `extra' coordinate produces
$$\int {\cal L}({\bf x},y) dy = \frac{\sigma}{(2\pi R)^{(N-1)/2}}
   \left(\prod_{i=1}^N \phi_{i k_i}\right)\delta_{\sum_i k_i,0},$$
with the sum taken over positive and negative $k$ values. Thus we deduce 
the effective coupling of $\sigma$ with the various $\phi$ in the lower 
dimension to be $g_{k_1..k_N} =\delta_{\sum_i k_i,0}/(2\pi R)^{(N-1)/2}.$
This means that the higher dimensional phase space can be written in the
more explicit form,
\begin{equation}
 \rho^D_{m_0\rightarrow m_1+\cdots+m_N}= \sum_{k_i}\frac{\delta_{\sum_i k_i,0}}
 {(2\pi R)^{N-1}} \rho^{D-1}_{m_0\rightarrow m_{1k_1}+\cdots+m_{Nk_N}},
\end{equation}
subject to energy-momentum conservation of course, which thus provides upper
bounds on the magnitudes of the running $k$-values.  

The second step is to 
take the limit as $(R,k)\rightarrow\infty$ and let $\mu_i=k_i/R$. We see that 
the connection (4) reduces to the continuous version,
\begin{equation}
 \rho^D_{m_0\rightarrow m_1+\cdots+m_N}\!=\!\left(\!\prod_i\!\int\!
 \frac{d\mu_i}{2\pi}\!\right)2\pi
 \delta(\sum_i \mu_i)\rho^{D-1}_{m_0\rightarrow\!m_{1\mu_1}\!+\cdots+m_{N\mu_N}};
 \, m^2_{i\mu}\!\equiv\! m_i^2\!+\!\mu^2,
\end{equation}
where again the range of $\mu$-values is restricted by the condition
$m_0\geq m_{1\mu_1}+\cdots+m_{N\mu_N}$. This last form is quite readily
understood as a consequence of writing the mass shell condition, $0= p^2-m^2 
= {\bf p}^2-(m^2 +\mu^2)$, where $\mu$ stands for the last momentum component 
$p_y$. The phase space integral over $p_y$ then produces the delta function
$\delta(\sum_i \mu_i)$, because the initial zero-mode has no dependence on 
$p_y$.

Equation (5) is the recurrence relation we were seeking, so now let us see how
it works out for those cases which we can tackle explicitly. Start with the
easiest case, $N=2$, where we know that
\begin{equation}
 \rho^{2\ell}_{m_0\rightarrow m_1+m_2}=\frac{\pi^{1-\ell}\Gamma(\ell-1)
 (\lambda(m_0^2,m_1^2,m_2^2))^{2\ell-3}}
  {2^{2\ell-1}m_0^{2\ell-2}\Gamma(2\ell-2)},
\end{equation}
involving the K\"allen function, $\lambda(a,b,c)\equiv\sqrt{a^2+b^2+c^2
-2ab-2bc-2ca}$. The recurrence relation (5) reduces to the prediction that
\begin{equation}
 \rho^{2\ell}_{m_0\rightarrow m_1+m_2}\!=\frac{1}{2\pi}\int d\mu\,\,
 \rho^{2\ell-1}_{m_0\rightarrow m_{1\mu}+m_{2\mu}}
 \theta(m_0-m_{1\mu}-m_{2\mu}), 
\end{equation}
and its verification relies upon the observation that
$$\lambda^2(m_0^2,m_{1\mu}^2,m_{2\mu}^2) =
  \lambda^2(m_0^2,m_1^2,m_2^2)-4m_0^2\mu^2,$$
plus the basic integral ($M^2\equiv \lambda(m_0^2,m_1^2,m_2^2)/2m_0$ and
$r\equiv 2\ell-3$ below)
\begin{eqnarray*}
\int d\mu\,\theta(m_0\!-\!m_{1\mu}\!-\!m_{2\mu})
       \lambda^r(m_0^2,m_{1\mu}^2,m_{2\mu}^2)\!\!&=&\!\!
 \int_0^{M^2}\frac{d\mu^2}{\sqrt{\mu^2}}\left(2m_0\sqrt{M^2-\mu^2}\right)^r;
  \nonumber \\
 &=&\!\!\frac{\lambda^{r+1}(m_0^2,m_1^2,m_2^2)\,\Gamma(\frac{r}{2}+1)\sqrt{\pi}}
            {2m_0\,\Gamma(\frac{r}{2}+\frac{3}{2})}.
\end{eqnarray*}

The recurrence relation becomes much more interesting for the three-body case,
\begin{equation}
 4\pi^2\rho^{2\ell}_{m_0\rightarrow m_1+m_2+m_3}=\int\!\!\!\int\!\!\!\int\!  
 d\mu_1\,d\mu_2\,d\mu_3\,\delta(\mu_1+\mu_2+\mu_3)
 \rho^{2\ell-1}_{m_0\rightarrow m_{1\mu_1}+m_{2\mu_2}+m_{3\mu_3}}
\end{equation}
upon recalling that even and odd dimensional phase space behave rather 
differently: odd $D$ leads to a Laurent series in the masses, while even $D$ 
generally leads to elliptic functions \cite{BDR,Alm,BBBB}; eq.(8) signifies 
that there exists an integral relation between elliptic functions and 
polynomials/poles. 
Moreover the nature of the integral displays {\em explicit} symmetry in the 
masses which is useful. Because of the $\delta$ function constraint and the 
fact that the $\mu$ run over positive and negative values, the rhs of (8) can 
be broken up into the sum of three terms:
$$ 4\pi^2\rho^{2\ell}_{m_0\rightarrow m_1+m_2+m_3}=2\int_0\!\!\int_0
   d\mu_1\,d\mu_2 
  \rho^{2\ell-1}_{m_0\rightarrow m_{1\mu_1}+m_{2\mu_2}+m_{3\mu_1+\mu_2}}
 \,{\rm~+~2~cyclic~perms.}$$
We may quickly check the truth of relation (8) for the test case $\ell=2$
when all particles are massless, since that limit of phase space simply 
yields
$$\rho^{4}_{m_0\rightarrow 0+0+0} =
  \lim_{\ell\rightarrow 2}\frac{(4\pi)^{1-2\ell}[\Gamma(\ell-1)]^3
   m_0^{4\ell-6}}{2\Gamma(3\ell-3)\Gamma(2\ell-2)} =\frac{m_0^2}{256\pi^3}$$
for the lhs. On the other hand for the rhs, substituting the 3-D result 
\cite{Raj},
$$\rho^3_{m_0\rightarrow m_1+m_2+m_3}=(m_0\!-\!m_1\!-\!m_2\!-\!m_3)
\theta(m_0\!-\!m_1\!-\!m_2\!-\!m_3)/16\pi m_0,$$
each of the three permutations produces the same answer and we obtain a perfect
check of the recurrence relation. However one learns something new in the 
massive case since a {\em new} symmetrical representation for 4-D phase space
emerges:
\begin{eqnarray}
 \rho^4_{m_0\rightarrow m_1+m_2+m_3}\!\!\!&=&\!\frac{1}{32\pi^3 m_0}
 \int_0\!\int_0 d\mu_1 d\mu_2\,
 (m_0\!-\!m_{1\mu_1}\!-\!m_{2\mu_2}\!-\!m_{3(\mu_1+\mu_2)}) \nonumber\\
 & & + {\rm 2~other~perms}.
\end{eqnarray}
One expects the right hand integral to produce elliptic functions, but the 
main virtue of (9) is that we get a pleasingly symmetrical sum of them which 
in principle ought to match an earlier form \cite{DD}, albeit obtained in a 
different manner. Although we have not yet succeeded in establishing the 
precise relation with the Jacobian zeta function form, we have performed a
series of {\em successful} numerical checks of (9), using typical mass values.

The same thing happens for larger $N$. For instance in 4-body decay there
arise four permutations with three of the $\mu_i$ having the same sign, 
opposite to the last one, plus six permutations where two pairs of $\mu_i$
have the same sign and opposite to the other pair. This would provide an
elegant symmetrical way of evaluating 4-body decay without resorting to
Almgren's nonsymmetrical way \cite{Alm} of pairing two bodies together and 
then summing over their pairwise mass sums.

One may of course continue in this vein and discuss spaces with the topology
$M^{D-n}\otimes T^n$, but one learns very little new by this ruse because the 
process just yields a set of angles $\theta_j=y_j/R_j$ and a set of mode 
numbers $k_j$ which collectively lead to $m_{\bf k}^2=m^2+\sum_j(k_j/R_j)^2$.
There is little gain in taking the limit as each $R_j\rightarrow\infty$
separately because we only care for the final result where none of the
radii is finite.

\section{Relations between $D$ and $D-2$ phase space}

Next we shall suppose that w are dealing with the direct topology $M^{D-2}
\otimes S^2$ so that the two extra angular coordinates are confined to the
surface of a 2-sphere having radius $R$; thus $\vec{y} = R
(\sin\theta\cos\varphi,\sin\theta\sin\varphi,\cos\theta)$, whereas previously
$y$ was identified with the circumference $R\theta$ rather than the radius 
vector. In such a situation expand the fields in spherical harmonics,
\begin{equation}
 \Phi({\bf x},\hat{y})=\sum_{j,k}\phi_{jk}({\bf x})Y_{jk}(\theta,\varphi)/R,
\end{equation}
before considering the limit of large $R$. The chosen factors ensure that
the lower-dimensional field modes $\phi$ are properly normalized:
\begin{equation}
 S_{\rm free}\!=\!\frac{1}{2}\int\!\! d^D\!x[(\partial_x\Phi)^2\!-\!m^2\Phi^2]
 =\sum_{j,k}\!\frac{1}{2}\int\!\!d^{D-2}\!{\bf x}[(\partial_{\bf x}\phi_{jk}^*)
 (\partial_{\bf x}\phi_{jk})\!-\!m_{j}^2\phi_{jk}^*\phi_{jk}],
\end{equation}
where $m_{j}^2=m^2+j(j+1)/R^2$. [In eq. (11) it is really meant that
$d^Dx = d^{D-2}{\bf x}R^2d^2\Omega$ and $(\partial \Phi)^2 = g^{ab}
\partial_a\Phi\partial_b\Phi$.] Again we note that $\phi_{jk}=\phi_{j-k}^*$
so the complex modes are found by just summing over positive $k$ and
discarding the factor of 1/2.

The two-body recurrence relation can be verified in its entirety, since the
effective coupling of a zero-mode mass $m_0$ field 
$\Sigma =\sigma/\sqrt{4\pi}R$ to two others yields a uniform amplitude, 
independently of the angular momentum eigenvalues $j,k$ as we see from
\begin{equation}
 \int d^Dx\, \Phi_1\Phi_2\Sigma= \sum_{j,k}
     \int d^{D-2}{\bf x}\,\,\sigma\Phi_{1j\,k}\Phi_{2j\,-k}/\sqrt{4\pi}R.
\end{equation}
The $(2j\!+\!1)$ degeneracy in $k$ leads to
\begin{equation} 
\rho^D_{m_0\rightarrow m_1+m_2}=\frac{1}{4\pi R^2}\sum_j(2j+1)
  \rho^{D-2}_{m_0\rightarrow m_{1j}+m_{2j}};\qquad 
  m_{ij}^2=m_i^2+\left(\frac{j}{R}\right)^2.
\end{equation}
In the limit $(R,j)\rightarrow \infty$, with $\mu=j/R$, this reduces to the 
continuum prediction,
\begin{equation}
 \rho^D_{m_0\rightarrow m_1+m_2}=\int_0 d\mu^2\,\,
 \rho^{D-2}_{m_0\rightarrow m_{1\mu}+m_{2\mu}}/4\pi,
\end{equation}
which is readily verified from the explicit result (6).

The three body case is altogether more fascinating. Here the effective 
interaction reduces to
\begin{eqnarray}
 \int d^Dx\, \Phi_1\Phi_2\Phi_3\Sigma&=&\frac{1}{\sqrt{4\pi}R^2}
  \sum_{j_1,k_1}\sum_{j_2,k_2}\sum_{j_3,k_3}\int d^{D-2}{\bf x}\,
  \sigma\Phi_{1j_1\,k_1}\Phi_{2j_2\,k_2}\Phi_{3j_3\,k_3}\cdot\nonumber \\
 & &\qquad\qquad\int d^2\Omega\,Y_{j_1k_1}Y_{j_2k_2}Y_{j_3k_3}.
\end{eqnarray}
To make progress, use the orthogonality property of spherical 
harmonics \cite{sphharm},
$$\int d^2\Omega\,Y_{j_1k_1}Y_{j_2k_2}Y_{j_3k_3}\!=\!
   \sqrt{\frac{(2j_1\!+\!1)(2j_2\!+\!1)(2j_3\!+\!1)}{4\pi}} 
   \left(\!\begin{array}{ccc} j_1\!&\!j_2\!&\!j_3\\0\!&\!0\!&\!0\end{array}
    \!\right)
 \left(\!\begin{array}{ccc} j_1\!&\!j_2\!&\!j_3\\k_1\!&\!k_2\!&\!k_3\end{array}
  \!\right).$$
This then specifies the magnitudes of the mode couplings. When evaluating
the sum over modes, apply the completeness relation of C-G coefficients,
$$\sum_{k_2,k_3}  
 \left(\!\begin{array}{ccc} j_1\!&\!j_2\!&\!j_3\\k_1\!&\!k_2\!&\!k_3\end{array}
 \!\right)
 \left(\!\begin{array}{ccc} j_4\!&\!j_2\!&\!j_3\\k_4\!&\!k_2\!&\!k_3\end{array}
 \!\right)
 =\frac{\delta_{j_1j_4}\delta_{k_1k_4}}{2j_1+1},$$
signifying
$$\sum_{k_1,k_2,k_3} \left(\!
 \begin{array}{ccc}j_1\!&\!j_2\!&\!j_3\\k_1\!&\!k_2\!&\!k_3\end{array}\!\right)^2
 =\sum_{k_1}\frac{1}{2j_1+1} = 1. $$
Therefore the summation over modes produces the discrete recurrence relation,
\begin{equation}
\rho^D_{m_0\rightarrow m_1+m_2+m_3}\!=\!\!\!\!\sum_{j_1,j_2,j_3}\!\!\!
 \frac{(2j_1\!+\!1)(2j_2\!+\!1)(2j_3\!+\!1)}{(4\pi R^2)^2}\left(\!\!
 \begin{array}{ccc}j_1\!&\!j_2\!&\!j_3\\ 0\!&\!0\!&\!0\end{array}\!\!\right)^2
 \!\!\rho^{D-2}_{m_0\rightarrow m_{1j_1}\!+m_{2j_2}\!+m_{3j_3}}
\end{equation}
whose continuum limit is of interest. To take $R\rightarrow\infty$, first note
that Wigner's 3-j symbol \cite{sphharm}
$$\hspace{-2in}\left(\!\!
 \begin{array}{ccc}j_1\!&\!j_2\!&\!j_3\\ 0\!&\!0\!&\!0\end{array}\!\!\right)
 \equiv (-1)^{-(j_1+j_2+j_3)/2}\times$$
$$\qquad\qquad\qquad\frac{ (\frac{j_1+j_2+j_3}{2})! \sqrt{(-j_1+j_2+j_3)!
   (j_1-j_2+j_3)!(j_1+j_2-j_3)!}} {\sqrt{(1+j_1+j_2+j_3)!}
 (\frac{-j_1+j_2+j_3}{2})!(\frac{j_1-j_2+j_3}{2})!(\frac{j_1+j_2-j_3}{2})!}.$$
Since we are concerned with the large $j$ limit, apply Stirling's formula,
$$ \frac{a!}{(a/2)!^2} \simeq \frac{2^{a+1/2}}{\sqrt{\pi a}},$$ 
to the previous expression. The square of the 3-j symbol then magically 
simplifies to the inverse area of a triangle having sides $j_1,j_2,j_3$:
\begin{equation}
 \left(\!\!
 \begin{array}{ccc}j_1\!&\!j_2\!&\!j_3\\ 0\!&\!0\!&\!0\end{array}\!\!\right)^2
 \simeq \frac{2\theta(\lambda_E)}{\pi\lambda_E(j_1^2,j_2^2,j_3^2)};
 \qquad \lambda_E^2\equiv -\lambda^2,
\end{equation}
which makes good sense, recalling the vector addition formula for
angular momenta. One finishes with the continuum result
\begin{equation}
 \rho^{D}_{m_0\rightarrow m_1+m_2+m_3}\!=\frac{1}{4\pi^3}
 \int_0\!\!\int_0\!\!\int_0 \!\!d\mu_1\,d\mu_2\,d\mu_3
 \frac{\mu_1\mu_2\mu_3}{\lambda_E(\mu_1^2,\mu_2^2,\mu_3^2)}
 \rho^{D-2}_{m_0\rightarrow m_{1\mu_1}+m_{2\mu_2}+m_{3\mu_3}}.
\end{equation}

This recurrence is very difficult to verify in general, especially for even
$D$. We can however make a fist of it for 5-D in the massless limit when
the check collapses to the veracity of
\begin{equation}
 \rho^5_{m_0\rightarrow 0+0+0}\!=\frac{1}{64\pi^4}
 \int_0\!\!\int_0\!\!\int_0 \!\!d\mu_1 d\mu_2 d\mu_3
 \frac{\mu_1\mu_2\mu_3\theta(\lambda_E)}{\lambda_E(\mu_1^2,\mu_2^2,\mu_3^2)}
 \left( 1\!-\frac{\mu_1+\mu_2+\mu_3}{m_0} \right).
\end{equation}
The lhs of (19) is known to equal $m_0^4/53760\pi^3$. To integrate the rhs, 
change variables according to $ \mu_1=w/2-u,\quad\mu_2=w/2-v,
\quad\mu_3=v+u\qquad {\rm so}$,
$$ \int_0\!\!\int_0\!\!\int_0\!\!d\mu_1d\mu_2d\mu_3\,X=2\int_0^{m_0}\!dw
  \int_0^{w/2}\!dv\int_0^{w/2-v}\!du\, X$$
may be carried out with $X\propto 
  \frac{(w/2-u)(w/2-v)(u+v)(m_0-w)}{\sqrt{uv[2w(u+v)-w^2]}}$. The result indeed 
reproduces the lhs. We have also verified the truth of (19) numerically.

The miraculous birth of the Euclidean K\"allen function $\lambda_E$ in (17) 
and (19) can be rendered less mysterious if we look upon this as the result of 
integrating over the last 2 components of momentum $\vec{\mu}$, corresponding 
to a `radial kernel'. Thus from the mathematical fact that
$$\left(\prod_{i=1}^3\int\frac{d^2\vec{\mu_i}}{(2\pi)^2}\right)(2\pi)^2\delta^2
 (\sum_{i=1}^3 \vec{\mu_i})F(|\vec{\mu_j}|) =
 \left(\prod_{i=1}^3 \int\frac{\mu_i d\mu_i d\theta_i}{(2\pi)^2} \right)\!
 F(\mu_j)\!\int\!\!d^2\vec{k}\,\,{\rm e}^{i\vec{k}\cdot\sum\vec{\mu}} $$
we find that this expression equals \cite{GH,MOSGrad}
$$\left(\!\prod_{i=1}^3\!\int\!\frac{\mu_id\mu_i}{2\pi}\!\!\right)\!\!F(\mu_j)
 2\pi\!\!\int_0^\infty\!\!\!\!\!\!J_0(k\mu_1\!)J_0(k\mu_2\!)J_0(k\mu_3\!)\,kdk
=\!\!\left(\!\prod_{i=1}^3\!\int\!\frac{\mu_id\mu_i}{2\pi}\!\!\right)\!
\!\frac{4\theta(\lambda_E)F(\mu_j)}{\lambda_{\!E}(\mu_1^2,\!\mu_2^2,\!\mu_3^2)}.$$
Moreover the extension to $N$-body phase space suggests itself immediately
via an $N$-fold kernel, associated with the integral
$2\pi\!\int_0^\infty\left(\prod_{i=1}^N J_0(k\mu_i)\right)\,kdk$
although this is not readily stated in terms of simple functions for $N>4$;
something geometrical associated with the lengths $\mu_i$ is certainly involved.
This kernel has to be folded over 
$\rho^{D-2}_{m_0\rightarrow m_{1\mu_1}+\cdots +m_{N\mu_N}}$
and integrated with respect to all $\mu_i d\mu_i$ to establish the recurrence.

\section{Relations between $D$ and $D-n$ phase space}

With the torus and 2-sphere thoroughly understood, it is natural to extend
the argument to coordinates confined to an $n$-sphere where the topology
is $M^{D-n}\otimes S^n$. Here we need to make use 
of hyperspherical harmonics \cite{Bateman} defined over $n$ angles. Associated 
with them are the generalized quadratic Casimir operator with eigenvalue 
$j(j+n-1)$ and $n-1$ angular momentum components (generically labelled 
${\bf k}$), producing a degeneracy of 
$h_{jn}=(2j+n-1).(j+n-2)!/j!(n-1)!$ Thus the free action may be
normalized according to eq. (11), where
$$\Phi({\bf x},\hat{y}) = \sum_{j,{\bf k}}\phi_{j,{\bf k}}({\bf x})\,
 Y_{j,{\bf k}}(\hat{y})/R^{n/2},$$
and we must sum over $(n-1)$ of the ${\bf k}$ labels. Now the squared mass
equals $m^2 + j(j+n-1)/R^2$ of course, because the last term corresponds to a 
hyperspherical Laplacian eigenvalue \cite{Bateman}.

Running through similar steps as before and skipping details, we arrive at
the two-body recurrence relation for finite $R$,
\begin{equation}
 \rho^D_{m_0\rightarrow m_1+m_2} = \sum_{j} h_{jn}
 \rho^{D-n}_{m_0\rightarrow m_{1j}+m_{2j}}/\Omega_nR^n,
\end{equation}
where $\Omega_n = 2\pi^{(n+1)/2}/\Gamma((n+1)/2)$ is the total solid angle
corresponding to $n$ angular coordinates. In the limit of large $R$ and 
therefore large $j$, since $h_{jn} \simeq 2j^{n-1}/\Gamma(n)$, we obtain the 
continuum limit,
\begin{equation}
 \rho^D_{m_0\rightarrow m_1+m_2}=\frac{\Gamma((n+1)/2)}{\pi^{(n+1)/2}\Gamma(n)}
 \int_0 d\mu \mu^{n-1} \rho^{D-n}_{m_0\rightarrow m_{1\mu}+m_{2\mu}}
\end{equation}
and this is readily checked via the explicit answer (6). Eqs. (7) and (14)
are particular cases of (21).

While we have succeeded in treating the two-body decay by this procedure, it
clearly becomes unwieldy and probably useless for $N>2$ and larger values
of $n$, since we would have to integrate over multiproducts of hyperspherical
harmonics, which are not exactly well-known, though some fancy generalisations 
of 3-j symbols and the like must exist. On the other hand one can make much 
better progress by regarding the recurrence as the result of
integrating over the last $n$ momenta $\vec{\mu}$:
\begin{equation}
 \rho^{D}_{m_0\rightarrow m_{1}+\cdots +m_{N}}=
 \left(\prod_{i=1}^N\int\frac{d^n\!\vec{\mu_i}}{(2\pi)^n}\right)
 \int d^n\vec{k}\,\,{\rm e}^{i\vec{k}.\sum_i\vec{\mu_i}} 
 \, \rho^{D-n}_{m_0\rightarrow m_{1\mu_1}+\cdots +m_{N\mu_N}}
\end{equation}
Now in general \cite{ADM},
$$\int d^n\!{\vec{\mu}} \exp(i\vec{k}.\vec{\mu})\,f(\mu)=
 \int_0^\infty (2\pi\mu)^{n/2}J_{n/2-1}(k\mu)\,f(\mu)\,d\mu/k^{n/2-1},$$
and $\,d^n\!\vec{k}=k^{n-1}dk\,\Omega_{n-1}=2k^{n-1}d\mu.\pi^{n/2}/\Gamma(n/2)$,
so (20) simplifies to
\begin{eqnarray}
 \rho^{D}_{m_0\rightarrow m_{1}+\cdots +m_{N}}&=&
 \left(\prod_{i=1}^N \int_0^\infty\frac{\mu_i^{n/2}d\mu_i}{(2\pi)^{n/2}}\right)
 \int_0^\infty \frac{2\pi^{n/2}k^{n-1}\,dk}{\Gamma(n/2)}\,
 \left(\prod_{i=1}^N \frac{J_{n/2-1}(k\mu_i)}{k^{n/2-1}}\right) \nonumber \\
 &&\qquad\qquad\times\rho^{D-n}_{m_0\rightarrow m_{1\mu_1}+\cdots +m_{N\mu_N}}.
\end{eqnarray}
As far as we are aware there is no amenable formula for the radial kernel: an 
integral over the product of $N$ Bessel functions of the first kind with 
different arguments for all $N$. However the cases $N=2$ and $N=3$ are known 
and for positive $a,b,c,$ read \cite{GH}:
$$\int_0^\infty J_\nu(ax)J_\nu(bx)\,xdx=2\delta(a^2-b^2),$$
$$\int_0^\infty J_\nu(ax)J_\nu(bx)J_\nu(cx)\,x^{1-\nu}dx =
 2\theta(\lambda_E)\lambda_E^{2\nu-1}/(8abc)^\nu\Gamma(\nu+1/2)\sqrt{\pi}.$$
For $N=4$ it is even known that \cite{MOSGrad}
$$\int_0^\infty J_0(ax)J_0(bx)J_0(cx)J_0(dx)\,xdx={\bf K}
  \left(\sqrt{abcd}/\Delta\right)/\pi^2\Delta,$$
where $16\Delta^2=(a+b+c-d)(b+c+d-a)(c+d+a-b)(d+a+b-c)$ is associated with the
maximal area of a (cocyclic) quadrilateral formed by the lengths $a,b,c,d$
and ${\bf K}$ is the complete elliptic integral of the first kind. Anyhow,
this means that we may at least write a simple closed form for the three-body
recurrence relation, when $n$ is arbitrary:
\begin{equation}
 \rho^{D}_{m_0\rightarrow m_1\!+m_2\!+m_3}\!=\!\!
 \left(\!\prod_{i=1}^3\!\int_0^\infty\!\frac{\mu_i d\mu_i}{(2\pi)^{n/2}}\!
 \right)\!\!\frac{2^{3-n/2}\theta(\lambda_E)\pi^{n/2-1}}
                    {\lambda_E^{3-n}(\mu^2,\mu_2^2,\mu_3^2)\Gamma(n\!-\!1)}
 \rho^{D-n}_{m_0\rightarrow m_{1\mu_1}\!+m_{2\mu_2}\!+m_{3\mu_3}}.
\end{equation}
An interesting case occurs when $N=n=3$, whereupon (24) reduces to
\begin{equation}
 \rho^{D}_{m_0\rightarrow m_1\!+m_2\!+m_3}\!=\!\int_0^\infty\!\!\int_0^\infty\!
 \!\int_0^\infty d\mu_1^2d\mu_2^2d\mu_3^2 \frac{\theta(\lambda_E)}{64\pi^4}
 \rho^{D-3}_{m_0\rightarrow m_{1\mu_1}\!+m_{2\mu_2}\!+m_{3\mu_3}}
\end{equation}
and because 3-D phase space is so simple \cite{Raj}, we obtain an intriguing 
representation for 6-D phase space on setting $D=6$. Specifically,
$$ \rho^6_{m_0\rightarrow m_1\!+m_2\!+m_3}\!\!=\!\int_0^\infty\!\!\!\!\!da
  \!\int_0^\infty\!\!\!\!\!db\!\int_0^\infty\!\!\!\!\!dc
  \frac{\theta(\lambda_E(a,\!b,\!c))}
  {64\pi^4}\frac{m_0\!-\!\sqrt{m_1^2\!+\!a}\!-\!\sqrt{m_2^2\!+\!b}
 \!-\!\sqrt{m_3^2\!+\!c}}{16\pi m_0}.
$$

With larger $N$, presumably the radial kernel on the right of (23) involves all 
lengths $\mu_i$ and something geometrically more complicated than $\lambda_E$, 
connected with the closed figure $\sum_i\vec{\mu_i}=0$. [Thus we expect it to 
vanish when any length exceeds the lengths of the other three, amongst other 
conditions. This is a very interesting topic worth future investigation.]

\section{Conclusions}
In this paper we have established the connection between relativistic phase 
space for different dimensions by two methods. They complement Almgren's 
connection between phase space
in the same dimension, but for different numbers of decay products. Our 
recurrence relations have practical utility when the difference in dimensions
$n$ is odd since it leads to elegant symmetrical representations of
$\rho$ for even $D$ involving elliptic functions.

Last but not least, one should observe that phase space is nothing but
the imaginary part of a sunset Feynman diagram. Since recurrence relations
between sunset diagrams differing in dimensionality by 2 have been found by
other methods \cite{DT}, it should be possible to rewrite our results that way, 
for even $n$ at any rate. We should also point out that as a rule those
relations do not connect even and odd $D$ because they are obtained by 
integration by parts from scalar particle Feynman graphs.

\section*{Acknowledgements}

We are pleased to acknowledge financial support from the Australian Research
Council under grant number A00000780. Comments and suggestions by 
A.I.~Davydychev are also much appreciated.


\end{document}